\documentclass[usenatbib]{mn2e}
\usepackage{graphicx}

\title[GSC~2314-0530: the shortest-period dMe system]
{GSC~2314-0530: the shortest-period eclipsing system with dMe components
\thanks{Based on the data obtained at Rozhen National Astronomical Observatory,
and the Northern Sky Variability Survey}}
\author[D. Dimitrov and D. Kjurkchieva]{Dinko P. Dimitrov$^{1}$\thanks{E-mail: dinko@astro.bas.bg; d.kyurkchieva@shu-bg.net} and Diana  P. Kjurkchieva$^{2}$\\
$^{1}$Institute of Astronomy, Bulgarian Academy of Sciences, 72 Tsarigradsko shossee str., 1784 Sofia, Bulgaria\\
$^{2}$Department of Physics, Shumen University, 115 Universitetska
str., 9700 Shumen, Bulgaria}

\begin{document}

\date{Accepted --. Received --; in original form --}

\pagerange{\pageref{firstpage}--\pageref{lastpage}} \pubyear{2010}

\maketitle

\label{firstpage}

\begin{abstract}
CCD photometric observations in $VRI$ colors and
spectroscopic observations of the newly discovered eclipsing binary
GSC~2314-0530 (NSVS~6550671) with dMe components and very short
period of $P=0.192636$ days are presented. The simultaneous
light-curve solution and radial velocity solution allows to
determine the global parameters of GSC~2314-0530: $T_{1}=3735$ K;
$T_{2}=3106$ K; $M_{1}=0.51$ M$_{\sun}$; $M_{2}=0.26$ M$_{\sun}$;
$R_{1}=0.55$ R$_{\sun}$; $R_{2}=0.29$ R$_{\sun}$; $L_{1}=0.053$
L$_{\sun}$; $L_{2}=0.007$ L$_{\sun}$; $i=72.5\degr$; $a=1.28$
R$_{\sun}$; $d=59$ pc. The chromospheric activity of its components
is revealed by strong emission in the H$\alpha$ line (with mean
$EW=5\ {\rm \AA}$) and observed several flares. Empirical relations
mass--$M_{\rm {bol}}$, mass--radius and mass--temperature are
derived on the basis of the parameters of known binaries with
low-mass dM components.
\end{abstract}


\section{Introduction}\label{sec:intro}

Although the M dwarfs are the most numerous stars in our Galaxy,
the mass, metalicity and age dependencies of their stellar
luminosities and radii are poorly calibrated. The reason is the
selection effect that plays against the detection of fainter and smaller
stars.

Less than 20 binaries with low-mass dM components have
empirically-determined masses, radii, luminosities and temperatures
(see Section \ref{sec:global}, Table \ref{tab:stars}). As a result
the mass-luminosity relation is determined by only a few low-mass
stars. This deficiency hindered the development of the models for
the cool dense atmospheres of the M dwarfs. It is
established that all available models underestimate the radii (by
around 10--15 per cent) and overestimate the temperatures (by
200--300 K) of short-period binaries with dM components
\citep{ribas03,maceroni04}.

The Northern Sky Variability Survey (NSVS) contains a great number
of photometric data \citep{wozniak04} that allows searching
of variable stars and determination of their periods and types of
variability. A multiparametric method for search for variable
objects in large datasets was tested on the NSVS
\citep{dimitrov09} and as a result many eclipsing stars
were discovered. One of them was GSC~2314-0530 $\equiv$ NSVS~6550671
($\alpha$=02$^{\rm h}20^{\rm m}50\fs9$, $\delta$=+$33\degr 20\arcmin
46\farcs6$).

\begin{figure}
 \centering
 \includegraphics[width=0.75\columnwidth]{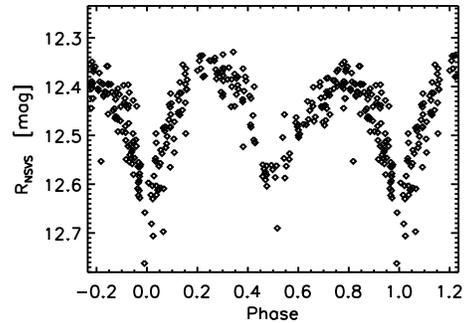}
\caption{NSVS photometry of GSC~2314-0530}
\label{fig:nsvs}
\end{figure}

On the base of the NSVS photometry obtained in 1999--2000 we
derived the ephemeris:

\begin{equation}\label{equ:nsvs}
HJD({\rmn {MinI}})=2451352.062 + 0.192637 \times E
\end{equation}
and built its light curve (Fig. \ref{fig:nsvs}).

We found that this star has been assigned also as
SWASP~J022050.85+332047.6 according to the SuperWASP photometric
survey \citep{pollacco06}. \citet{norton07} reported its coincidence
with the {\it ROSAT} X-ray source 1RXS~J022050.7+332049.

Initially GSC~2314-0530 attracted our interest by its short orbital
period because there were only several systems with non-degenerate
components and periods below the short-period limit of 0.22
days \citep{rucinski07}: GSC~1387-0475 with $P=0.217811$ d
\citep{rucinski07,rucinski08}, ASAS~J071829-0336.7 with $P=0.211249$
d \citep{pribulla09}, the star V34 in the globular cluster 47 Tuc
with $P=0.2155$ d \citep{weldrake04} and BW3~V38 with orbital period
$P=0.1984$ d \citep{maceroni97,maceroni04}.

When we established that the components of GSC~2314-0530 were dM
stars our interest increased and we undertook intensive photometric
and spectral observations in order to determine its global
parameters and to add a new information for the dM stars as well as
for the short-period binaries.

\begin{figure}
 \centering
 \includegraphics[width=0.75\columnwidth]{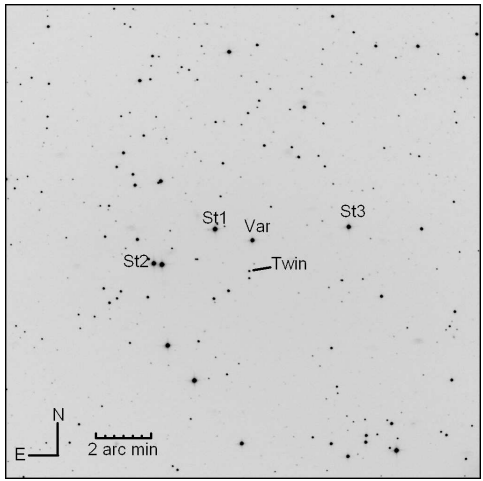}
 \caption{Observed field around GSC~2314-0530}
 \label{fig:chart}
\end{figure}

\section{Observations and data reduction}\label{sec:observation}

\subsection{New photometry}

The CCD photometry of GSC~2314-0530 in $VRI$ bands was carried out
at Rozhen National Astronomical Observatory with the 2-m RCC
telescope equipped with VersArray CCD camera (1300 $\times$ 1340
pixels, 20  $\mu$m pixel, field of 5.25 $\times$ 5.35 arcmin) as
well as with the 60-cm Cassegrain telescope using the FLI PL09000
CCD camera (3056 $\times$ 3056 pixels, 12  $\mu$m pixel, field of
17.1 $\times$ 17.1 arcmin). The average photometric precision per
data point was 0.005 -- 0.008 mag for the 60-cm telescope and 0.002
-- 0.003 mag for the 2-m telescope. Table \ref{tab:log1} presents
the journal of our photometric observations.

It should be noted that the observations on 2009 December 30 are
synchronous in the $VRI$ colors.

\begin{table*}
\begin{minipage}[t]{\textwidth}
\caption{Journal of the photometric observations}
\label{tab:log1}
\centering
\renewcommand{\footnoterule}{}
\begin{tabular}{ccccrrc}
\hline \hline
Date   &  HJD(start)  & Phases & Filter   & Exp. [s] &  N  & Telescope \\
\hline
2009 July 25 & 2455038.482662 & 0.725 -- 1.298 & $R$ & 120 & 126 & 60-cm \\
2009 July 26 & 2455039.484907 & 0.927 -- 0.491 & $R$ & 120 &  54 & 60-cm \\
2009 July 27 & 2455040.468322 & 0.032 -- 0.693 & $R$ & 120 &  83 & 60-cm \\
2009 July 28 & 2455041.501250 & 0.395 -- 0.881 & $R$ & 120 &  62 & 60-cm \\
2009 Oct. 21 & 2455126.412740 & 0.210 -- 1.294 & $V$ &  15 & 737 &  2-m  \\
2009 Nov. 13 & 2455149.178102 & 0.393 -- 1.389 & $I$ &  10 & 850 &  2-m  \\
2009 Nov. 13 & 2455149.375822 & 0.419 -- 1.391 & $R$ &  10 & 835 &  2-m  \\
2009 Nov. 20 & 2455156.324421 & 0.489 -- 0.521 & $B$ & 120 &   3 & 60-cm \\
2009 Nov. 20 & 2455156.325521 & 0.495 -- 1.862 & $V$ &  60 & 183 & 60-cm \\
2009 Nov. 20 & 2455156.326088 & 0.498 -- 0.529 & $R$ &  30 &   3 & 60-cm \\
2009 Nov. 20 & 2455156.326493 & 0.500 -- 0.531 & $I$ &  30 &   3 & 60-cm \\
2009 Dec. 30 & 2455196.225416 & 0.610 -- 1.785 & $V$ & 120 &  65 & 60-cm \\
2009 Dec. 30 & 2455196.226516 & 0.616 -- 1.791 & $R$ &  60 &  65 & 60-cm \\
2009 Dec. 30 & 2455196.227256 & 0.619 -- 1.810 & $I$ &  60 &  65 & 60-cm \\
\hline
\end{tabular}
\end{minipage}
\end{table*}

Standard stars of \citet{landolt92} and standard fields of
\citet{stetson00} were used for transition from the instrumental
system of each telescope to standard photometric system.

The standard IDL procedures (adapted from DAOPHOT) were used for
reduction of the photometric data. The standard stars were chosen on
the basis of the method of \citet{everett01} and Table
\ref{tab:colors} presents their colors. The values of $J-K$ are from
the catalogue NOMAD \citep{zacharias05} while the values of other
parameters are our estimations.

The field of the variable and standard stars is shown in Fig.
\ref{fig:chart}.

\begin{table*}
\begin{minipage}[t]{\textwidth}
\caption{Colors and proper motion of the variable star and standard
stars} \label{tab:colors} \centering
\renewcommand{\footnoterule}{}
\begin{tabular}{l c c c c c c c c}
\hline \hline
       &     ID       & $V$   & $B-V$ & $V-R$ & $V-I$ & $J-K$ & pmRA   & pmDE   \\
       & GSC/USNO-B1  & [mag] & [mag] & [mag] & [mag] & [mag] & [mas yr$^{-1}$] & [mas yr$^{-1}$] \\
\hline
Var    & 2314-0530    & 13.36 &  1.18 &  0.88 &  2.38 &  0.87 &  144.0 & -112.0 \\
St1    & 2314-1784    & 12.12 &  0.30 &  0.25 &  0.57 &  0.29 & -000.8 & -008.3 \\
St2    & 2314-1378    & 12.24 &  0.29 &  0.24 &  0.58 &  0.34 & -000.1 & -001.6 \\
St3    & 2314-1655    & 12.40 &  0.22 &  0.20 &  0.46 &  0.27 &  005.5 & -004.0 \\
Twin   & 1233-0046425 & 16.91 &  1.41 &  1.03 &  3.02 &  0.87 &  140.0 & -112.0 \\
\hline
\end{tabular}
\end{minipage}
\end{table*}

Table \ref{tab:photometry} presents a sample of our
photometric data (the full table is available in the online version
of the article, see Supporting Information).

\begin{table}
\begin{minipage}[t]{\columnwidth}
\caption{BVRI photometry of GSC~2314-0530}
\label{tab:photometry} \centering
\renewcommand{\footnoterule}{}
\begin{tabular}{l c c}
\hline \hline
  HJD &  Magnitude & Filter \\
\hline
2455156.329669 & 14.8530 & B \\
2455156.332679 & 14.8490 & B \\
2455156.335689 & 14.8300 & B \\
2455126.417320 & 13.3610 & V \\
2455126.418512 & 13.3618 & V \\
2455126.419890 & 13.3619 & V \\
2455126.420167 & 13.3624 & V \\
2455126.420700 & 13.3547 & V \\
2455126.420978 & 13.3583 & V \\
2455126.421128 & 13.3609 & V \\
\hline
\end{tabular}
\end{minipage}
\end{table}

\begin{figure*}
 \centering
 \includegraphics[width=0.7\textwidth]{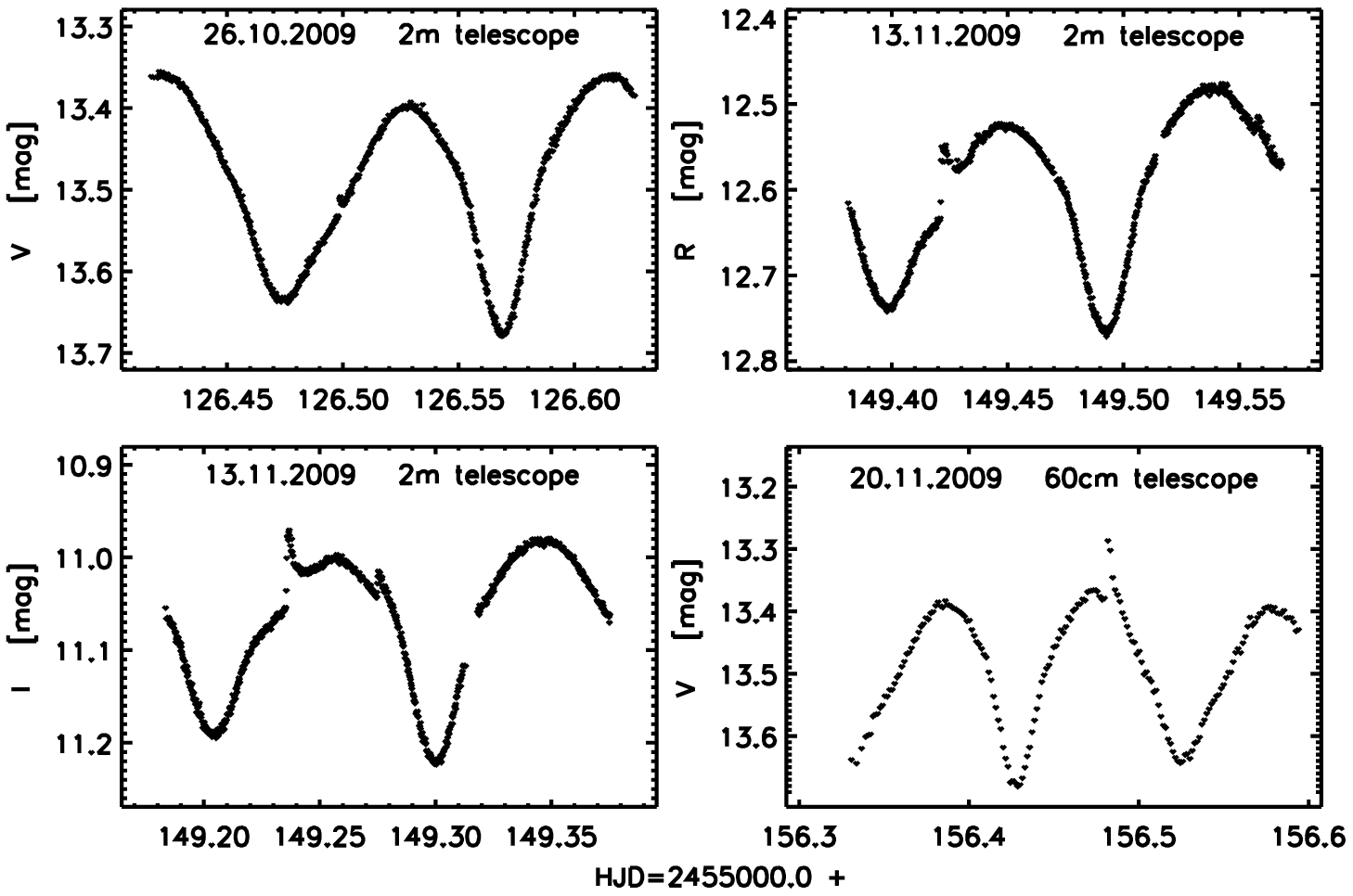}
\caption{The photometric observations of GSC~2314-0530 from 2009
October 26, 2009 November 13 and 2009 November 20} \label{fig:m/HJD}
\end{figure*}

Some of our photometric runs covering well the orbital cycle are
presented in Fig. \ref{fig:m/HJD}.

The Fourier analysis of all our photometric data performed by the
software PERIOD-04 \citep{lenz05} leads to the ephemeris:

\begin{equation}\label{equ:rozhen}
 HJD({\rm MinI})=2451352.061633 + 0.1926359\times E .
\end{equation}
The new-obtained period value is almost the same as that of the
ephemeris (\ref{equ:nsvs}) of the NSVS data that means that the
orbital period of GSC~2314-0530 is stable.

The color indices of our target (Table \ref{tab:colors}) lead to M 
spectral type of the binary. Taking into account the almost equal 
eclipse depths of the light curve, i.e. the close temperatures of the 
components, as well as the short orbital period of the system, we may 
conclude that the two components of GSC~2314-0530 are dM stars.

The value of the obtained period is below the short-period
limit and reveals that our target is the shortest-period binary with
dM components.

\begin{figure}
 \centering
 \includegraphics[width=0.9\columnwidth]{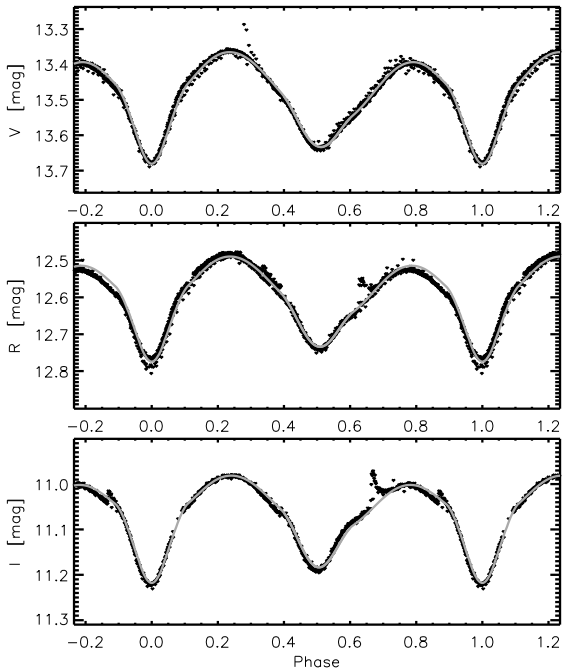}
 \caption{The folded $V$, $R$, $I$ light curves of GSC~2314-0530 and their fits}
 \label{fig:folded}
\end{figure}

Figure \ref{fig:folded} shows the folded light curves from all our photometric
data phased according to the ephemeris (\ref{equ:rozhen}).

\subsection{Spectroscopy}

We obtained 26 spectra of GSC~2314-0530 with resolution $0.19\ {\rm{\AA}}$/pixel
during November -- December 2009 covering spectral
range of $200\ {\rm{\AA}}$ around the H$\alpha$ line. We used a
CCD Photometrics AT200 camera with the SITe SI003AB 1024 $\times$
1024 pixels chip mounted on the Coude spectrograph (grating
B$\&$L632/14.7$\degr$) on the 2-m RCC telescope at Rozhen.

The exposure time was 15 min during 2009 November 26 and 20 min during
2009 December 31 and 2010 January 01. All stellar integrations were
alternated with Th-Ar comparison source exposures for wavelength
calibration. The bias frames and flat-field integrations were
obtained at the beginning and at the end of the night. The mean
S/N ratio for our observations was around 24, i.e. acceptable for
radial velocity determination. Table \ref{tab:radvel} presents the
journal of our spectral observations.

\begin{figure}
 \centering
 \includegraphics[width=\columnwidth]{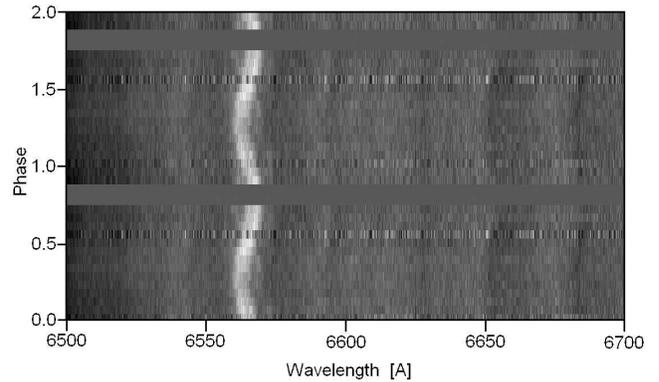}
 \caption{The orbital variability of the spectra of GSC~2314-0530 from 2009 November 26}
 \label{fig:Ha-synoptic}
\end{figure}

\begin{figure}
 \centering
 \includegraphics[width=0.75\columnwidth]{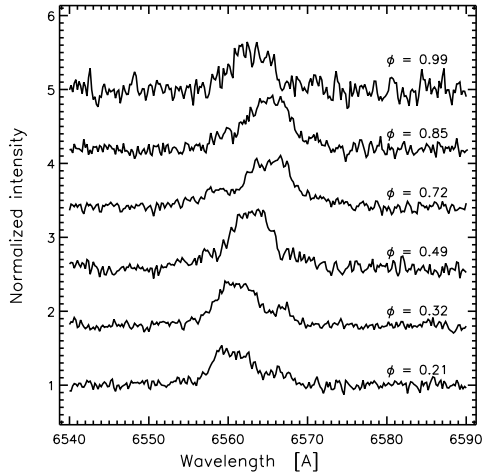}
 \caption{The H$\alpha$ profiles at some phases}
 \label{fig:Ha-phases}
\end{figure}

The reduction of the spectra was performed using IRAF packages by
bias subtraction, flat fielding, cosmic ray removal,
one-dimensional spectrum extraction and wavelength calibration.
Figure \ref{fig:Ha-synoptic} illustrates the orbital variability
of the star spectra while Figure \ref{fig:Ha-phases} presents the
one-dimensional H$\alpha$ profiles at some orbital phases.

\begin{figure}
 \centering
 \includegraphics[width=0.75\columnwidth]{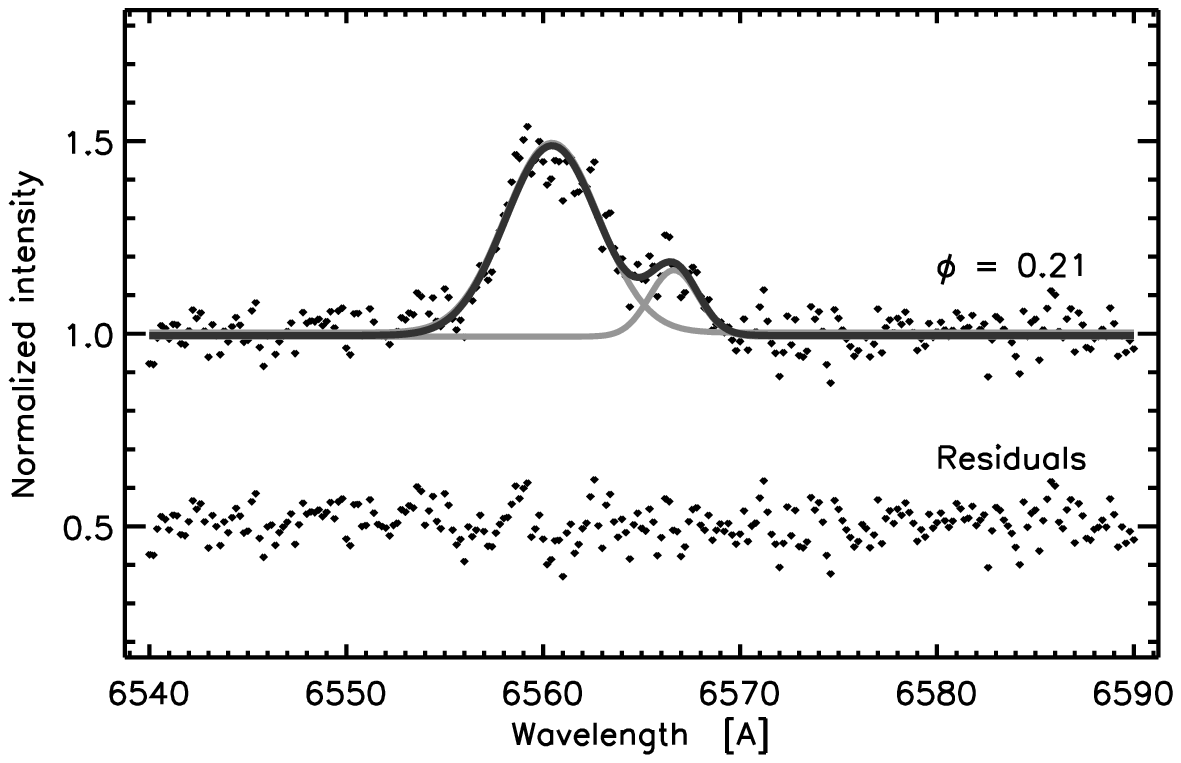}
 \caption{The two Gaussians (gray lines) reproducing the H$\alpha$ line (dots) of the two stellar components,
and their sum (black line) fitting the total H$\alpha$ profile of GSC~2314-0530.}
 \label{fig:Ha-fit}
\end{figure}

\section{Analysis of the spectral data}\label{sec:analysis_sp}

The obtained spectra of GSC~2314-0530 show wide emission
H$\alpha$ lines implying high rotational velocities as well as
absorption TiO bands at $6569\ {\rm{\AA}}$ and $6651\ {\rm{\AA}}$
(Fig. \ref{fig:Ha-synoptic}). These spectral features suggest a dMe
classification of GSC~2314-0530.

The spectral contribution of the secondary component is visible only
in the H$\alpha$ line (Fig. \ref{fig:Ha-synoptic}). That is why we
determined the radial velocities of the two stellar components by
fitting the H$\alpha$ lines at each phase with Gaussians
(Fig. \ref{fig:Ha-fit}).

\begin{figure}
\centering
\includegraphics[width=0.75\columnwidth]{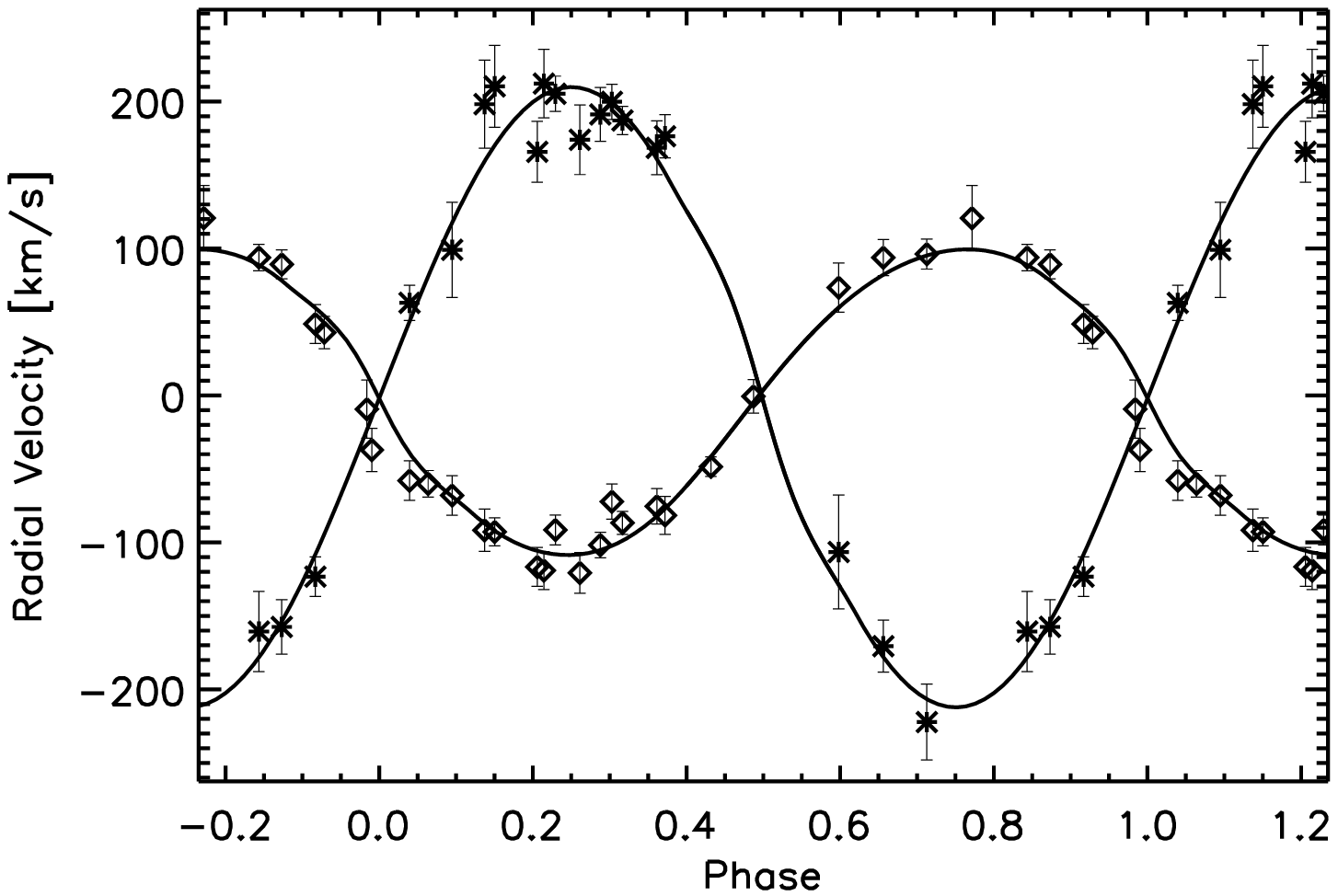}
\caption{Radial velocities of the two components of GSC~2314-0530 
(the sizes of the error bars correspond to 3$\sigma$) and their fits 
by the code PHOEBE \citep{prsa05}.} \label{fig:RV}
\end{figure}

Table \ref{tab:radvel} and Figure \ref{fig:RV} present the radial
velocities of the stellar components of GSC~2314-0530. Their fit
corresponds to values $K_{1}=V_{1}\sin i = 109.7\pm3.2$ km s$^{-1}$,
$K_{2} = V_{2}\sin i = 211.3\pm5.8$ km s$^{-1}$ and $V_{0}\sin i
=-1.2\pm5.7$ km s$^{-1}$. They lead to mass ratio $q=0.519\pm0.029$
and binary separation $a \sin i=1.22\pm0.04$ R$_{\sun}$.

\begin{table*}
\begin{minipage}[t]{\textwidth}
\caption{Journal of the spectral observations and parameters of the
H$\alpha$ lines} \label{tab:radvel} \centering
\renewcommand{\footnoterule}{}
\begin{tabular}{lcccrcrcc}
\hline \hline
No &     HJD       & S/N  & phase & \multicolumn{2}{c}{$RV_1$} & \multicolumn{2}{c}{$RV_2$} & $EW_{\rm{total}}$ \\
&     &      &      &  \multicolumn{2}{c}{[km s$^{-1}$]} & \multicolumn{2}{c}{[km s$^{-1}$]} & [$\rm{\AA}$] \\
\hline
01 & 2455162.375293 & 23 & 0.88 &  89.2 & $\pm3.3$ & -157.5 & $\pm6.2$ & 4.97  \\
02 & 2455162.385975 & 19 & 0.93 &  42.8 & $\pm3.7$ &        &          & 4.56  \\
03 & 2455162.396657 & 22 & 0.99 &  -9.3 & $\pm6.6$ &        &          & 3.61  \\
04 & 2455162.407335 & 21 & 0.04 & -57.9 & $\pm4.5$ &   63.0 & $\pm4.0$ & 6.64  \\
05 & 2455162.418012 & 27 & 0.10 & -68.0 & $\pm4.5$ &   99.1 & $\pm10.8$& 4.34  \\
06 & 2455162.428689 & 28 & 0.15 & -92.8 & $\pm3.2$ &  210.3 & $\pm9.3$ & 4.04   \\
07 & 2455162.439367 & 28 & 0.21 &-116.6 & $\pm4.4$ &  165.8 & $\pm6.9$ & 3.53   \\
08 & 2455162.450048 & 29 & 0.27 &-120.7 & $\pm4.6$ &  174.0 & $\pm7.9$ & 3.62  \\
09 & 2455162.460726 & 28 & 0.32 & -86.7 & $\pm2.6$ &  187.1 & $\pm3.2$ & 4.60  \\
10 & 2455162.471405 & 30 & 0.38 & -81.6 & $\pm4.3$ &  176.4 & $\pm4.9$ & 4.49  \\
11 & 2455162.482903 & 25 & 0.44 & -48.4 & $\pm2.2$ &        &          & 5.47   \\
12 & 2455162.493582 & 26 & 0.49 &  -0.6 & $\pm3.8$ &        &          & 5.55   \\
13 & 2455162.514939 & 25 & 0.60 &  73.4 & $\pm5.6$ & -106.5 & $\pm12.9$& 5.34   \\
14 & 2455162.526143 & 24 & 0.66 &  93.8 & $\pm4.1$ & -170.6 & $\pm5.9$ & 4.94  \\
15 & 2455162.537036 & 25 & 0.72 &  96.2 & $\pm3.4$ & -222.2 & $\pm8.6$ & 6.18    \\
16 & 2455197.222839 & 18 & 0.78 & 120.7 & $\pm7.4$ &        &          & 5.92    \\
17 & 2455197.236667 & 29 & 0.85 &  93.9 & $\pm3.0$ & -160.6 & $\pm9.1$ & 5.07   \\
18 & 2455197.250817 & 27 & 0.92 &  48.6 & $\pm4.4$ & -123.3 & $\pm4.5$ & 5.34   \\
19 & 2455197.264965 & 28 & 0.99 & -37.1 & $\pm4.9$ &        &          & 5.59  \\
20 & 2455197.279111 & 27 & 0.07 & -60.1 & $\pm3.0$ &        &          & 5.67   \\
21 & 2455197.293257 & 29 & 0.14 & -91.7 & $\pm4.8$ &  198.2 & $\pm10.0$& 4.16   \\
22 & 2455197.308131 & 29 & 0.22 &-119.1 & $\pm4.3$ &  212.2 & $\pm7.8$ & 4.44  \\
23 & 2455197.322278 & 30 & 0.29 &-101.8 & $\pm2.9$ &  191.2 & $\pm6.1$ & 3.76   \\
24 & 2455197.336431 & 28 & 0.37 & -75.4 & $\pm4.0$ &  168.5 & $\pm6.1$ & 3.65   \\
25 & 2455198.274197 & 25 & 0.23 & -91.5 & $\pm3.4$ &  205.3 & $\pm4.3$ & 6.04   \\
26 & 2455198.288351 & 25 & 0.31 & -72.2 & $\pm4.0$ &  199.8 & $\pm4.0$ & 6.29    \\
\hline
\end{tabular}
\end{minipage}
\end{table*}

\section{Analysis of the photometric data}\label{sec:analysis_ph}

The qualitative analysis of the new photometric data
(Fig.~\ref{fig:folded}) leads to several conclusions.

\begin{figure}
 \centering
 \includegraphics[width=0.75\columnwidth]{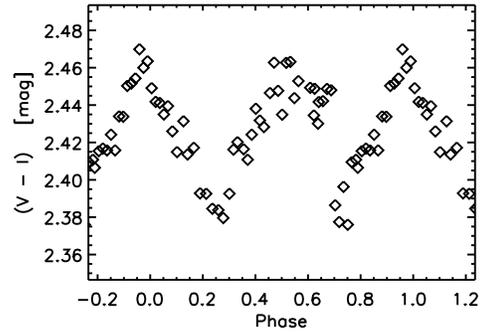}
 \caption{$V-I$ light curve of GSC~2314-0530 from synchronous $VRI$ observation
with 60-cm telescope}
 \label{fig:V-I}
\end{figure}

\begin{description}
\item[(1)] The Min~I is deeper than Min~II. This means that the
secondary's temperature is lower than the primary's temperature.
\item[(2)] The light maxima are not equal. This O'Connell effect
implies presence of surface temperature spot(s).
\item[(3)] The Max~I appears at the expected phase 0.25 while the
phase of Max~II is around 0.78. As a result the second half of the
light curves is quite distorted. Similar asymmetry is visible also
on the NSVS light curve (Fig.~\ref{fig:nsvs}) of the star almost 10
years earlier, i.e. this distortion is possibly permanent.
We noted that the shape of the light curve of GSC~2314-0530 at phase
range 0.5--0.8 resembles at some degree that of the cataclysmic
stars with their peculiar standstills causing delay of the light
increasing after the light minimum.
\item[(4)] The $V-I$ light curve of GSC~2314-0530 (Fig.~\ref{fig:V-I})
clearly reveals that the system becomes redder after the two
eclipses and bluer after the two quadratures. The phases of the extrema of the
 $V-I$ light curve have around 0.05 phase delays in respect to those of the light curves
\textbf{except for} the second maximum of $V-I$ which delay is more than 0.10.
\item[(5)] We observed several flares of GSC~2314-0530 (Fig.~\ref{fig:m/HJD})
resembling those of UV Ceti stars (see more in Section \ref{sec:activity}).
\end{description}

In order to determine the global parameters of GSC~2314-0530 we
modeled our $VRI$ folded curves simultaneously using the software
PHOEBE \citep{prsa05} by the following procedure.

\begin{description}
\item[(a)] We fixed the mass ratio $q=0.519$ from our
radial velocity solution.
\item[(b)] The obtained components of the heliocentric space velocity 
$U=-23$ km/s, $V=-44$ km/s and $W=-12$ km/s allow us to assume solar metalicity 
for the emission of GSC~2314-0530 \citep{leggett00}.
\item[(c)] We adopted coefficients of gravity brightening $g_1=g_2=0.32$
and reflection $A_1=A_2=0.5$ (appropriate for late stars) while the
limb-darkening coefficients for each star and each color were taken
from the tables of \citet{vanhamme93}.
\item[(d)] Taking into account that $E(V-I)=0.03$ mag in the GSC~2314-0530'
direction \citep{schlegel98} we obtained its de-reddened color index
$(V-I)_{0}=2.35$ mag. According to table 2 of \citet{vandenberg03}
this out-of-eclipse color index corresponds to mean temperature of
the binary $T_{\rm m}=3560$ K.
\end{description}

It should be noted that the index $B-V=1.18$ mag of GSC~2314-0530
corresponds to mean temperature around 4400 K, i.e. 840 K higher
than that obtained by the $V-I$ index. This is a new confirmation of
the conclusion that the majority of the dMe stars have $B-V$ colors too blue
for their $V-I$ colors \citep{stauffer86}. Our result also shows
that the temperature difference obtained by the two color indices
($V-I$ and $B-V$) is higher than 200--300 K \citep{ribas03,maceroni04}
and can reach 800 K.

\begin{description}
\item[(e)] At the first stage we fixed $T_{1}=3700$ K (taking into account
that the temperature of the primary component $T_{1}$ is higher than
$T_{\rm m}$) and varied the secondary' temperature $T_{2}$, the orbital
inclination $i$ and the potentials $\Omega_{1,2}$. In order to
reproduce the O'Connell effect and light curve distortions we had to
add two cool spots on the primary's surface and to vary their
parameters: longitude $\lambda$, latitude $\beta$, angular size $\alpha$ and temperature $T_{\rm {sp}}$.
\end{description}

Moreover, in order to get a good simultaneous fit for the three
colors $VRI$ by the same stellar and spot parameters we added a
third light $L_3$ which contributes differently to the
different colors. We consider the last supposition as artificial
step to compensate the peculiar energy distribution of the dM stars
that appear especially faint in the $V$ band probably to the big TiO
absorption as well as to the big contribution of the spots.

\begin{description}
\item[(f)] After getting a good fit of our $VRI$ photometric data we began
to vary also the primary's temperature. As a result we obtained the
best light curve solution which parameters are given in Table
\ref{tab:lightsolution}. The respective synthetic $VRI$ light curves
are shown in Fig. \ref{fig:folded} as gray lines. They coincide very
well with the observational data at all phases except for the flares.
\end{description}

\begin{table}
\begin{minipage}[t]{\columnwidth}
\caption{Best light curve solution from {\sc phoebe}}
\label{tab:lightsolution} \centering
\renewcommand{\footnoterule}{}
\begin{tabular}{clll}
\hline \hline
\multicolumn{2}{c}{Parameter} & \multicolumn{2}{c}{Value} \\
\hline
$   i   $         & [\degr] & 72.5 & $\pm0.1$  \\
$  T_1  $         &  [K]  & 3735  & $\pm10$    \\
$  T_2  $         &  [K]  & 3106  & $\pm10$    \\
$\Omega_1$        &       & 2.944 & $\pm0.002$ \\
$\Omega_2$        &       & 3.545 & $\pm0.009$ \\
$\lambda_{\rm {Sp1}}$  & [\degr] & 147   & $\pm5$     \\
$\beta_{\rm {Sp1}}$   & [\degr] & 70    & $\pm10$    \\
$\alpha_{\rm {Sp1}}$     & [\degr] & 20    & $\pm1$     \\
$T_{\rm {Sp1}}$   & [K]   & 3175  &    $\pm50$        \\
$\lambda_{\rm {Sp2}}$  & [\degr] & 195   &  $\pm5$    \\
$\beta_{\rm {Sp2}}$   & [\degr] & 75    &  $\pm10$   \\
$\alpha_{\rm {Sp2}}$     & [\degr] & 8     &  $\pm1$    \\
$T_{\rm {Sp2}}$   & [K]   & 3175  &      $\pm50$      \\
$L_3$(V)  &       & 0.171 & $\pm0.003$ \\
$L_3$(R)  &       & 0.222 & $\pm0.002$ \\
$L_3$(I)  &       & 0.298 & $\pm0.002$ \\
\hline
\end{tabular}
\end{minipage}
\end{table}

\begin{figure}
\centering
\includegraphics[width=0.95\columnwidth]{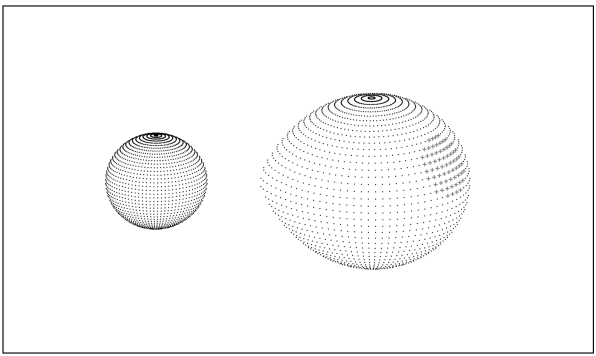}
\caption{3D model of GSC~2314-0530 at phase 0.75}
\label{fig:3D-config}
\end{figure}

The obtained potentials correspond to relative mean stellar radii
$r_{1}=0.431$ and $r_{2}=0.228$ revealing that the primary component
almost fills-in its Roche lobe (Fig. \ref{fig:3D-config}).

\section{Global parameters of GSC~2314-0530}\label{sec:global}

Using the photometric value of the orbital inclination
$i=72.5\degr$ we determined consecutively the following global
parameters of GSC~2314-0530:

\begin{description}
\item[(a)] orbital velocities of the two components  $V_{1}=115.1\pm3.4$ km s$^{-1}$,
$V_{2}=221.6\pm6.1$ km s$^{-1}$;
\item[(b)] orbital separation $a=1.28\pm0.04$ R$_{\sun}$;
\item[(c)] masses of the components $M_{1}=0.51\pm0.02$
M$_{\sun}$ and $M_{2}=0.26\pm0.02$ M$_{\sun}$;
\item[(d)] absolute (mean) radii of the components $R_{1}=0.55\pm0.01$ R$_{\sun}$
and $R_{2}=0.29\pm0.01$ R$_{\sun}$;
\item[(e)] surface gravity $\log g_1 = 4.68 $ and $\log g_2 = 4.95 $;
\item[(f)] stellar luminosities $L_{1}=0.053\pm0.002$ L$_{\sun}$
and $L_{2}=0.0070\pm0.0006$ L$_{\sun}$;
\item[(g)] bolometric absolute magnitudes of the components (using $M^{\sun}_{\rm {bol}}=4.72$)
$M_{\rm {bol1}}=7.91\pm0.04$ mag and $M_{\rm {bol2}}=10.11\pm0.09$
mag as well as bolometric absolute magnitude of the binary $M_{\rm
{bol}}(\rm total)=7.77\pm0.05$ mag;
\item[(h)] absolute $V$ magnitude of the binary
$M_{V}(\rm total)=9.5\pm0.05$ mag (using $BC_{V}=-1.73$ corresponding
to $T_{\rm m}$ from table 2 of \citealt{vandenberg03});
\item[(i)] distance to the binary $d=59\pm2$ pc.
\end{description}

It should be noted that while the masses and radii of the components
were directly determined, their temperatures and absolute magnitudes
required external calibrations which are poorly known for the late
stars.

We calculated the equatorial velocities of the components by
measuring the rotation broadening of their H$\alpha$ lines (using
$i=72.5\degr$). The obtained values $V_{\rm {rot1}}=145\pm15$ km
s$^{-1}$ and $V_{\rm {rot2}}=69\pm15$ km s$^{-1}$ reveal that the
components of GSC~2314-0530 are quite fast rotators (see Table
\ref{tab:activity}). Thus our target confirms the conclusion of
\citet{stauffer86} that the stars with larger velocities have
centrally peaked H$\alpha$ emission while the slower rotators have
centrally reversed profiles as well as the conclusion of
\citet{worden81} that stars with centrally peaked H$\alpha$ emission
profiles belong to short-period binaries.

\begin{table*}
 \begin{minipage}[t]{\textwidth}
\caption{Parameters of binaries with low-mass dM components}
\label{tab:stars}
\renewcommand{\footnoterule}{}\centering
\begin{tabular}{l c c c c c c c c r c r r r}
\hline \hline
Name            & $P$  & $T$  & $M$  & $R$  & $L$   & $i$& $q$  &$V-I$&$M_{\mathrm {bol}}$& $a$& $d$ &Type& Ref. \\
                &[d]&[K]&[M$_{\sun}$]&[R$_{\sun}$]&[L$_{\sun}$]&[\degr]& &[mag]&[mag]&[R$_{\sun}$]&[pc]&    & \\
\hline
CU Cnc=GJ 2069A & 2.77 & 3160 & 0.43 & 0.43 & 0.016 & 86 & 0.92 &2.80 & 9.19   &0.92&12.8& D  & (1)   \\
                &      & 3125 & 0.40 & 0.39 & 0.013 &    &      &     & 9.45   &    &    &    &\\
2MASS J01542930+0053266& 2.64 & 3700 & 0.66 & 0.64 & 0.069 & 86 & 0.95 &     & 7.62   & 8.70& 623& D  & (2) \\
                &      & 3300 & 0.62 & 0.61 & 0.039 &    &      &     & 8.24   &    &    &    &\\
NSVS~06507557   & 0.51 & 3960 & 0.65 & 0.60 & 0.079 & 83 & 0.42 & 2.13& 7.48   &2.65&111 & D  & (3) \\
                &      & 3360 & 0.28 & 0.44 & 0.022 &    &      &     & 8.86   &    &    &    &\\
NSVS~07394765   & 2.26 & 3170 & 0.56 & 0.58 & 0.030 & 84 & 1.16 &     & 8.52   &2.60 &    &  D & (4)  \\
                &      & 3860 & 0.65 & 0.69 & 0.095 &    &      &     & 7.27   &    &    &    & \\
NSVS~07453183   & 0.37 & 3340 & 0.68 & 0.72 & 0.060 & 89 & 1.07 & 1.40& 7.77   &7.75&    & D  & (4) \\
                &      & 3570 & 0.73 & 0.79 & 0.090 &    &      &     & 7.33   &    &    &    & \\
UNSW-TR-2       & 2.11 & 3870 & 0.53 & 0.64 & 0.082 & 83 & 0.95 &     & 7.43   &7.05& 169& D  & (5)\\
                &      & 3845 & 0.51 & 0.61 & 0.073 &    &      &     & 7.56   &    &    &    & \\
CM Dra          & 1.27 & 3150 & 0.23 & 0.25 & 0.005 & 90 & 0.93 &     &10.47   &3.75&    & D  & (6)\\
                &      & 3125 & 0.21 & 0.23 & 0.004 &    &      &     &10.71   &    &    &    &\\
TrES~HerO-07621 & 1.12 & 3500 & 0.49 & 0.45 & 0.027 & 83 & 0.95 &     & 8.64   &2.25& 118&  D &(7) \\
                &      & 3400 & 0.49 & 0.45 & 0.024 &    &      &     & 8.77   &    &    &    & \\
YY Gem          & 0.81 & 3820 & 0.60 & 0.62 & 0.070 & 86 & 1.00 & 1.92& 7.57   &3.87&    &  D &(8) \\
                &      & 3820 & 0.60 & 0.62 & 0.070 &    &      &     & 7.57   &    &    &    &\\
GJ 3226         & 0.77 & 3313 & 0.38 & 0.37 & 0.016 & 83 & 0.75 & 2.73& 9.20   &3.08& 42 &  D &(9) \\
                &      & 3247 & 0.28 & 0.32 & 0.009 &    &      &     & 9.83   &    &    &    & \\
2MASS~04463285+1901432 & 0.62 & 3320 & 0.47 & 0.56 & 0.034 & 81 & 0.41 & 2.59& 8.39   &2.66& 540& D  &(10)\\
                &      & 2910 & 0.19 & 0.21 & 0.003 &    &      &     &11.03   &    &    &    &\\
V405 And        & 0.496& 4050 & 0.49 & 0.78 & 0.147 & 66 & 0.98 &     & 6.80   &2.25&    & D  & (11) \\
                &      & 3000 & 0.21 & 0.23 & 0.004 &    &      &     &10.71   &    &    &    &\\
GU Boo          & 0.49 & 3920 & 0.61 & 0.62 & 0.082 & 88 & 0.98 & 1.90& 7.43   &2.79& 100& D  &(12) \\
                &      & 3810 & 0.60 & 0.62 & 0.073 &    &      &     & 7.60   &    &    &    &\\
SDSS~MEB-1      & 0.41 & 3320 & 0.27 & 0.27 & 0.008 & 85 & 0.98 &     & 9.96   &1.85&    & D  &(13)\\
                &      & 3300 & 0.24 & 0.25 & 0.007 &    &      &     &10.11   &    &    &    &\\
NSVS~01031772   & 0.37 & 3615 & 0.54 & 0.53 & 0.043 & 86 & 0.92 &     & 8.08   &2.20 & 40 & D  &(14)\\
                &      & 3513 & 0.50 & 0.51 & 0.036 &    &      &     & 8.27   &    &    &    &    \\
OGLE~BW3 V38    & 0.198& 3500 & 0.44 & 0.51 & 0.035 & 86 & 0.95 &2.45 & 8.39   &1.35& 400& SD &(15) \\
                &      & 3450 & 0.41 & 0.44 & 0.025 &    &      &     & 8.78   &    &    &    &\\
GSC~2314-0530   & 0.192& 3735 & 0.51 & 0.55 & 0.053 & 72 & 0.52 & 2.34& 7.91   &1.28& 59 & SD &(16)\\
                &      & 3106 & 0.26 & 0.29 & 0.007 &    &      &     &10.11   &    &    &    &  \\
\hline
\end{tabular}
\end{minipage}
References: (1) Ribas 2003, Delfosse et al. 1999; (2) Becker et al.
2008; (3) Cakirli $\&$ Ibanoglu 2009; (4) Coughlin $\&$ Shaw 2007;
(5) Young et al. 2006; (6) Metcalfe et al. 1996; (7) Creevey et al.
2005; (8) Bopp 1974, Torres $\&$ Ribas 2002; (9) Irwin et al. 2009;
(10) Hebb et al. 2006; (11) Vida et al. 2008; (12) Lopez-Morales
$\&$ Ribas 2005; (13) Blake et al. 2008; (14) Lopez-Morales et al.
(2006); (15) Maceroni $\&$ Montalban (2004); (16) this paper
\end{table*}

Some of the determined global parameters of GSC~2314-0530 together
with those of the other known binaries with low-mass dM components
are given in Table \ref{tab:stars} which columns are: star name;
period $P$ in days; temperatures $T$ of the components; masses $M$,
radii $R$ and luminosities $L$ of the components in solar units;
orbital inclination $i$ in degrees; mass ratio $q$; color index
$V-I$ of the binary; bolometric absolute magnitudes $M_{\rm {bol}}$
of the components; orbital separation $a$ in solar radii; distance
$d$ in pc; type of binary configuration (D -- detached, SD --
semidetached); references.

Figure \ref{fig:relations} shows the empirical diagrams mass-$M_{\rm
{bol}}$, mass-radius and mass-temperature for the low-mass stars
from Table \ref{tab:stars} (total number 34). They occupy relative
narrow bands on these diagrams. This means that the luminosities,
radii and temperatures of these stars depend on their masses. These
statistical relations can be described by the following formulas:

\begin{figure}
\centering
\includegraphics[width=0.85\columnwidth]{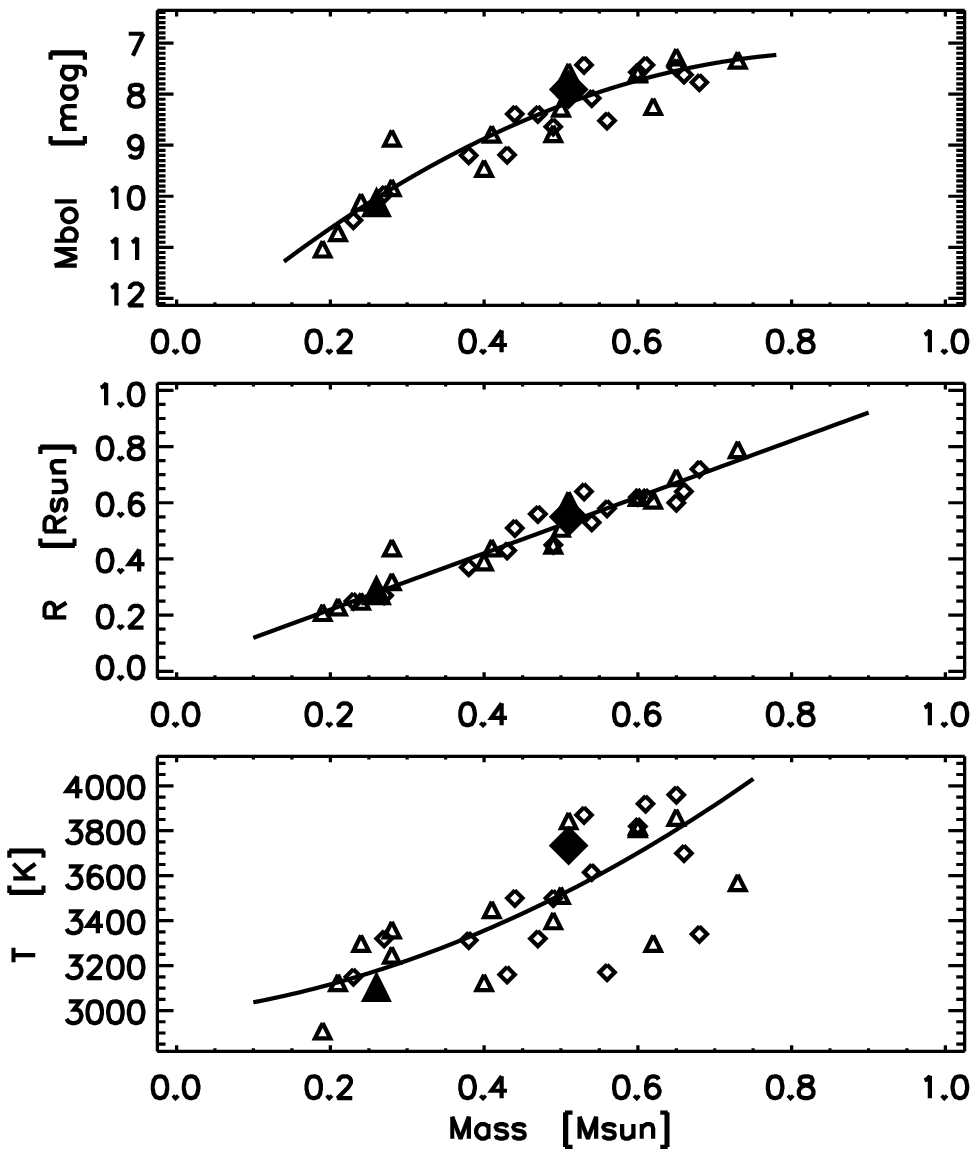}
\caption{Empirical relations mass-$M_{\rm {bol}}$, mass-radius and
mass-temperature for the low-mass stars from Table \ref {tab:stars}
and their fits. The locus of the primary stars are signed by
diamonds, those of the secondaries -- by triangles, while those of
our star -- by large filled symbols.}
 \label{fig:relations}
\end{figure}

\begin{eqnarray}
M_{\rm {bol}} & = & 13.0 - 13.4 \times M + 7.7 \times M^2 \\
\nonumber   R & = & 0.019 + 1.002 \times M \\
\nonumber   T & = & 2983 + 396 \times M + 1333 \times M^2
\end{eqnarray}

We assume that the bigger scatter of the mass-temperature diagram is
due mainly to the weakly established calibration $T$/$(V-I)$ for the
late low-mass stars. Moreover, some star temperatures probably have
been determined without taking into account the reddening.

\section{Activity of GSC~2314-0530}\label{sec:activity}

The manifestations of stellar activity as H$\alpha$ emission, spots,
flares, etc., are consequences of magnetic fields. It is assumed
that the fully-convective late stars have strong, long-lasting,
magnetic field.

According to \citet{mullan01} the larger radii and lower
temperatures of dM stars can be explained by the presence of strong
magnetic fields and their activity is at the saturation limit.
Perhaps the significant spot coverage decreases the photospheric
temperature which the star compensates by increasing its radius to
conserve the total radiative flux.

\subsection{Surface spots}

The photospheric activity of the late stars is demonstrated mainly
by O'Connell effect and distorted light curves. They can be
reproduced by surface temperature inhomogeneities
(spots). It is reasonable to assume existence of cool spots by
analogy with our Sun. Usually they are put on the primary star
although the same effect can be reached by spots on the secondary
but then the spots should be larger and/or cooler.
There are also fits of the light curves of late binaries with
bright spots \citep{torres02,maceroni94}. These are
interpreted by uniform distribution of dark spots covering
however most of stellar surface except for a spot-free area, i.e.
"bright spots" represent the true photosphere.

The light curves of all binaries with low-mass dM components from
Table \ref{tab:stars} are distorted and they have been reproduced by
large cool spots which angular radii reach up to 80$\degr$.

The distorted light curves of GSC~2314-0530 were reproduced by two
cool spots on the primary component (see their parameters in Table
\ref{tab:lightsolution}) covering 3.5 per cent of its surface. The
fact that the shape of the light curve distortions of GSC~2314-0530
remains the same almost 10 years means that the main (larger) spot
visible at phase 0.6 presents long-lived active region on the
primary surface.

\subsection{H$\alpha$ emission}

The $EW$ of the H$\alpha$ line is an useful indicator of
chromospheric activity for M dwarfs because those stars are
much brighter at 6500 ${\rm{\AA}}$ than at 3900 ${\rm{\AA}}$.
\citet{stauffer86} divided dM into 4 subsets ordered by
chromospheric activity. The least chromospheric active dM have weak
H$\alpha$ absorption line. As the chromosphere increases the $EW$ of
the H$\alpha$ absorption first increases, then decreases and finally
H$\alpha$ goes into the emission.

Table \ref{tab:radvel} presents the orbital variations of the $EW$
of the total H$\alpha$ emission of GSC~2314-0530. Although it seemed
to change irregularly in the range 3.6-6.6 ${\rm{\AA}}$ during the
cycle we noted a trend of the $EW$ to be smaller around the
first quadrature than around the second quadrature. The exceptions
from this trend are the big $EW$ values of the only two spectra from
2010 January 01 at phases 0.23 and 0.31. They may due to flare
event. Such a supposition is reasonable because two of the observed
flares are around the first quadrature (see Table \ref{tab:flares}).

The foregoing trend of the H$\alpha$ emission is opposite to that of
total light of GSC~2314-0530 that is bigger at the first quadrature
than at the second one. Such an anti-correlation is typical for the
chromospherically active stars of types RS CVn and BY Dra.

\begin{table}
\begin{minipage}[t]{\columnwidth}
\caption{Activity of low-mass dM stars} \label{tab:activity}
\centering
\renewcommand{\footnoterule}{}
\begin{tabular}{lrrrr}
\hline \hline
Star &$V_{\rm {rot1}}$&$V_{\rm {rot2}}$&$EW$  & Flares  \\
 & [km s$^{-1}$] & [km s$^{-1}$] & [${\rm{\AA}}$] & \\
\hline
CM Dra        &    &    &  em    & Y      \\
CU Cnc        &    &    & 4      &        \\
V405 And      &    &    &        & Y      \\
GU Boo        & 64 & 64 & 1.7    &        \\
YY Gem        & 37 & 37 & 2      & Y      \\
NSVS~06507557 & 59 & 43 &[-3,+2] &        \\
NSVS~01031772 & 72 & 70 &        &        \\
BW3 V38       & 131&113 &5.4     &        \\
GJ3236        & 25 & 19 &        &        \\
GSC~2314-0530 & 145& 69 & $\leq$6.6& Y      \\
\hline
\end{tabular}
\end{minipage}
\end{table}

Table \ref{tab:activity} presents the $EW$ of the H$\alpha$ emission
of some binaries with low-mass dM components from Table
\ref{tab:stars} at normal state (out of flare). The comparison
reveals the strong H$\alpha$ emission of GSC~2314-0530. This result
is not surprising taking into account the low temperature and fast
rotation of its components.

The mean value $EW=5\ {\rm{\AA}}$ of the H$\alpha$ emission
of GSC~2314-0530 is considerably smaller than that of the accreting
pre-main-sequence dMe stars which H$\alpha$ emission has $EW > 10\
{\rm{\AA}}$.

\subsection{Flares}

Flare activity is typical for the late stars. The last column of Table
\ref{tab:activity} shows those stars from our Table \ref{tab:stars}
in which some flares have been registered (signed by ``Y'').

During our observational runs we were witnesses of six flares of
GSC~2314-0530 that revealed its high flare activity. The amplitudes $A$
and durations $\tau$ of the observed flares are given in Table
\ref{tab:flares}.

It should be noted that 3 of the observed 6 flares occurred around
the phase of maximum visibility 0.6 of the large, stable spot (Sp1).
This implies correlation between the two signs of stellar
activity: spots and flares. Both of them are appearances of
the long-lived active area on the primary star.

Besides the optical flares there is information about X-flares of
GSC~2314-0530 \citep{fuhrmeister03}.

\begin{table}
\begin{minipage}[t]{\columnwidth}
\caption{Observed flares of GSC~2314-0530}
\label{tab:flares}
\centering
\renewcommand{\footnoterule}{}
\begin{tabular}{lrrcrr}
\hline \hline
Date    &HJD$_{\rm {max}}$  & Phase & Filter & $A$ & $\tau$ \\
           & $2455000 +$   &       &        &     [mag]     & [min]  \\
\hline
2009 Oct. 26 & 126.49373 & 0.61 &   $V$  &    0.022      &      4      \\
2009 Nov. 13 & 149.23146 & 0.64 &   $I$  &    0.085      &     22      \\
2009 Nov. 13 & 149.26995 & 0.84 &   $I$  &    0.027      &     13      \\
2009 Nov. 13 & 149.41788 & 0.61 &   $R$  &    0.085      &     19      \\
2009 Nov. 13 & 149.55281 & 0.31 &   $R$  &    0.015      &      9      \\
2009 Nov. 20 & 156.48180 & 0.31 &   $V$  &    0.092      &     25      \\
\hline
\end{tabular}
\end{minipage}
\end{table}

\subsection{Angular momentum}

The small orbital angular momentum is characteristic feature of all
short-period systems ranging from CVs to CB that seem to be old,
being at later stages of the angular momentum loss evolution as a
result of the period decrease.

We calculated the orbital angular momentum of the target by the
expression \citep{popper77}

\begin{equation}\label{equ:am}
J_{\rm {rel}}=M_{1}M_{2} \left( \frac {P}{M_{1}+M_{2}} \right)^{1/3}
\end{equation}
where $P$ is in days and $M_{i}$ are in solar units.

The obtained value $\log J_{\rm {rel}}=-1.01$ of GSC~2314-0530 is
considerably smaller than those of the RS CVn binaries and detached
systems which have $\log J_{\rm {rel}}\geq +0.08$. The orbital
angular momentum of GSC~2314-0530 is smaller even than those of the
contact systems which have $\log J_{\rm {rel}} \geq -0.5$. It is
bigger only than those of the short-period CVs of SU UMa type.

The small orbital angular momentum of GSC~2314-0530 implies
existence of past episode of angular momentum loss during the binary
evolution. It means also that GSC~2314-0530 is not pre-MS object.
This conclusion is supported by the values of $\log g$ of its
components.

\subsection{X-ray emission}

The X-ray emission of the stellar coronae are directly
related to the presence of magnetic fields and consequently gives
information about the efficiency of the stellar dynamo.

\citet{rucinski84} established that the X-ray luminosity
decreased for later M stars while the ratio $L_{X}/L_{bol}$ did not
change significantly from M0 to M6. As a result he proposed the
ratio $L_{X}/L_{bol}$ as most relevant measure of activity of M
dwarfs. \citet{vilhu87} found that the upper boundary of
$L_{X}/L_{bol}$ for late M stars is $\sim 10^{-3}$ .

Besides all indicators of stellar activity in the optical (surface
inhomogeneities, emission lines, flares) the star GSC~2314-0530
shows also X-ray emission (it is identified as {\it ROSAT} X-ray
source 1RXS~J022050.7+332049) and X-ray flares.

On the basis of the measured X-ray flux $F_{X}=4.266 \times
10^{-13}$ ergs cm$^{-2}$ $s^{-1}$ of GSC~2314-0530 at quiescence
\citep{voges99,schmitt95} and derived distance 59 pc we calculated
its X-ray luminosity $L_{X}=1.68 \times 10^{29}$ ergs s$^{-1}$. This
value is at the upper boundary $\log L_{X}\approx $ 29 for dM stars
\citep{rosner81,caillault86}. The value $f_{X}/f_{bol}=
L_{X}/L_{bol}=0.7 \times 10^{-3}$ of GSC~2314-0530 is almost at the
upper boundary of this ratio and considerably bigger than those of
the M dwarfs studied by \citet{rucinski84} and \citet{caillault86}.

It is known that the activity and angular momentum loss tend to be
saturated at high-rotation rates \citep{vilhu87}. Due to its short
period and high activity GSC~2314-0530 is perhaps an example of such
saturation.

\section{Is GSC~2314-0530 alone?}\label{sec:twin}

Our observed field (Fig. \ref{fig:chart}) contains the weak star
USNO-B1 1233-0046425. We called it Twin due to the same tangential
shift as our target star GSC~2314-0530. Table \ref{tab:colors}
presents the proper motion and the colors of Twin according to the
catalogue NOMAD. USNO-B1~1233-0046425 has $V-I=3.02$ corresponding
to temperature less than 3200 K.

We suspect that our "twins" may form visual binary. The angular
distance between them of 61 arcsec corresponds to linear separation
around 3500 au for distance of 59 pc. Such a supposition is
reasonable because it is known that the short-period close binaries
often are triple systems \citep{pribulla06}. Particularly, the
object TrES~Her0-07621 from our Table \ref{tab:stars} has a red
stellar neighbor at a distance 8 arcsec with close proper motion
\citep{creevey05}.

The check of the supposition if Twin is physical companion of
GSC~2314-0530 needs astrometric observations of the "twins".

\section{Conclusions}\label{sec:conclusion}

The analysis of our photometric and spectral observations of the
newly discovered eclipsing binary GSC~2314-0530 allows us to derive
the following conclusions:

(1) This star is the shortest-period binary with dM components which
period is below the short-period limit.

(2) By simultaneous radial velocity solution and light curve
solution we determined the global parameters of GSC~2314-0530:
inclination $i=72.5\degr$; orbital separation $a=1.28$ R$_{\sun}$;
masses $M_{1}=0.51$ M$_{\sun}$ and $M_{2}=0.26$ M$_{\sun}$; radii
$R_{1}=0.55$ R$_{\sun}$ and $R_{2}=0.29$ R$_{\sun}$; temperatures
$T_{1}=3735$ K and $T_{2}=3106$ K; luminosities $L_{1}=0.053$
L$_{\sun}$ and $L_{2}=0.007$ L$_{\sun}$; distance $d=59$ pc.

(3) We derived empirical relations mass--$M_{\rm {bol}}$,
mass--radius and mass--temperature on the basis of the parameters of
known binaries with low-mass dM components.

(4) The distorted light curve of GSC~2314-0530 were reproduced by
two cool spots on the primary component. The next sign of the
activity of GSC~2314-0530 is the strong H$\alpha$ emission of its
components. Moreover we registered 6 flares of GSC~2314-0530. Half
of them occurred at the phases of maximum visibility of the larger
stable cool spot on the primary.

The analysis of all appearances of magnetic activity revealed
existence of long-lived active area on the primary of GSC~2314-0530.
The high activity of the target is natural consequence of the fast
rotation and low temperatures of its components.

Our study of the newly discovered short-period eclipsing binary
GSC~2314-0530 presents a next small step toward understanding dMe
stars and adds a new information to the poor statistic of the
low-mass dM stars. Recently they became especially interesting as
appropriate targets for planet searches due to the relative larger
transit depths.

\section*{Acknowledgments}

The research was supported partly by funds of projects DO~02-362 of
the Bulgarian Scientific Foundation. This research make use of the
SIMBAD and Vizier databases, operated at CDS, Strasbourg, France,
and NASA's Astrophysics Data System Abstract Service. The
authors are very grateful to the anonymous referee for the valuable
notes and advices.

\label{lastpage}

\end{document}